\documentclass[aps,twocolumn,prd,superscriptaddress,preprintnumbers,showpacs]{revtex4-1}
\usepackage{graphicx}
\usepackage{bm}
\usepackage{amsmath}
\usepackage{amssymb}
\usepackage{amsthm}

\newcommand{\be}{\begin{equation}}
\newcommand{\ee}{\end{equation}}
\newcommand{\bea}{\begin{eqnarray}}
\newcommand{\eea}{\end{eqnarray}}
\newcommand{\bdm}{\begin{displaymath}}
\newcommand{\edm}{\end{displaymath}}
\newcommand{\beas}{\begin{eqnarray*}}
\newcommand{\eeas}{\end{eqnarray*}}

\begin{document}
\title{Current and future constraints on Bekenstein-type models for varying couplings}

\author{A. C. O. Leite}
\email[]{Ana.Leite@astro.up.pt}
\affiliation{Centro de Astrof\'{\i}sica, Universidade do Porto, Rua das Estrelas, 4150-762 Porto, Portugal}
\affiliation{Instituto de Astrof\'{\i}sica e Ci\^encias do Espa\c co, CAUP, Rua das Estrelas, 4150-762 Porto, Portugal}
\affiliation{Faculdade de Ci\^encias, Universidade do Porto, Rua do Campo Alegre, 4150-007 Porto, Portugal}
\author{C. J. A. P. Martins}
\email[]{Carlos.Martins@astro.up.pt}
\affiliation{Centro de Astrof\'{\i}sica, Universidade do Porto, Rua das Estrelas, 4150-762 Porto, Portugal}
\affiliation{Instituto de Astrof\'{\i}sica e Ci\^encias do Espa\c co, CAUP, Rua das Estrelas, 4150-762 Porto, Portugal}
\date{11 May 2016}

\begin{abstract}
Astrophysical tests of the stability of dimensionless fundamental couplings, such as the fine-structure constant $\alpha$ and the proton-to-electron mass ratio $\mu$, are an optimal probe of new physics. There is a growing interest in these tests, following indications of possible spacetime variations at the few parts per million level. Here we make use of the latest astrophysical measurements, combined with background cosmological observations, to obtain improved constraints on Bekenstein-type models for the evolution of both couplings. These are arguably the simplest models allowing for $\alpha$ and $\mu$ variations, and are characterized by a single free dimensionless parameter, $\zeta$, describing the coupling of the underlying dynamical degree of freedom to the electromagnetic sector. In the former case we find that this parameter is constrained to be $|\zeta_\alpha|<4.8\times10^{-6}$ (improving previous constraints by a factor of 6), while in the latter (which we quantitatively compare to astrophysical measurements for the first time)  we find  $\zeta_\mu=(2.7\pm3.1)\times10^{-7}$; both of these are at the $99.7\%$ confidence level. For $\zeta_\alpha$ this constraint is about 20 times stronger than the one obtained from local Weak Equivalence Principle tests, while for $\zeta_\mu$ it is about 2 orders of magnitude weaker. We also discuss the improvements on these constraints to be expected from the forthcoming ESPRESSO and ELT-HIRES spectrographs, conservatively finding a factor around 5 for the former and around 50 for the latter.
\end{abstract}

\maketitle

\section{\label{intro}Introduction} 

Now that we know (as a result of experiments at the Large Hadron Collider) that fundamental scalar fields are among Nature's building blocks \cite{ATLAS,CMS}, an obvious follow-up question is whether such scalar fields also play an role on cosmological scales. Among the astrophysical probes of such dynamical degrees of freedom, tests of the stability of Nature's dimensionless fundamental couplings are the most direct and model-independent. Whenever dynamical scalar fields are present, one naturally expects them to couple to the rest of the model, unless a yet-unknown symmetry suppresses these couplings \cite{Carroll}. In particular, a coupling to the electromagnetic sector will lead to spacetime variations of the fine-structure constant, which unavoidably imply a violation of the Einstein Equivalence Principle---see \cite{Uzan,GRG} for recent reviews.

Astrophysical tests of the stability of fundamental couplings are an extremely active area of observational research. The deep conceptual importance of carrying out these tests has been complemented by recent (even if somewhat controversial) evidence for such a variation \cite{Webb}, coming from high-resolution optical/UV spectrosopic measurements of absorption systems along the line of sight of bright quasars. A significant effort is being put into independently confirming this result, including a dedicated Large Program with the Very Large Telescope's UVES spectrograph \cite{LP1,LP2,LP3}. Improving these tests is also a flagship science case for forthcoming facilities such as ESPRESSO \cite{ESPRESSO} and ELT-HIRES \cite{HIRES}; a roadmap for these future tests is discussed in  \cite{GRG}.

Meanwhile, an extensive range of theoretical models including varying couplings have been studied (see \cite{Uzan} for an overview). Among these is a class of phenomenological models first suggested by Bekenstein \cite{Bekenstein} where, by construction, the dynamical degree of freedom responsible for the varying coupling  has a negligible effect on the cosmological dynamics. This includes the Sandvik-Barrow-Magueijo model for a varying fine-structure constant $\alpha$ (which stems from a varying electric charge)  \cite{SBM} and the more recent Barrow-Magueijo model for a varying proton-to-electron mass ratio $\mu$ (which stems from a varying electron mass) \cite{BM}.

The original works provided some qualitative constraints on the models \cite{SBM,BM}, an in particular kept the values of the cosmological parameters fixed at their best-fit values. This last assumption was also used in a more recent analysis of the $\alpha$ model \cite{Leal}. In this work we provide a more quantitative analysis, relying both on cosmological background data, astrophysical measurements of $\alpha$ and $\mu$ in the approximate redshift range $0<z<4.3$ and atomic clock laboratory bounds on the current drift rates of both quantities.

These models are characterized by a single phenomenological dimensionless parameter, $\zeta$, describing the strength of the coupling of the dynamical scalar degree of freedom to the electromagnetic sector, and we therefore obtain improved constraints on this parameter for both models. This same parameter also determines the amount of Weak Equivalence Principle (WEP) violation in these models, and we also compare our astrophysical constraints on $\zeta$ to those that follow from local WEP tests.

We note that in our analysis we will be considering the simplest version of both models. While extensions of the $\alpha$ model  with additional (functional) degrees of freedom have been suggested \cite{extend1,extend2}, the quantity and quality of the available data (and the fact that no strong evidence for a nonzero coupling $\zeta$ currently exists) motivate us to restrict ourselves to the simplest scenarios. Still, future observations will certainly allow the testing of broader scenarios. Here we also provide an illustration of this, quantifying the improvements on constraints on the simplest $\alpha$ scenario expected from the ESPRESSO spectrograph.

\section{\label{modela}The varying $\alpha$ model}

\subsection{The model and relevant data}

The Bekenstein-Sandvik-Barrow-Magueijo model was introduced in \cite{SBM}, drawing on earlier work from Bekenstein \cite{Bekenstein}. It is a model where the variation of $\alpha\equiv e^2/{\hbar c}$ is due to a varying electric charge, while other parameters are assumed to remain constant. We succinctly review its relevant features, referring the reader to the original work for more detailed derivations as well as some additional motivation. Conceptually one can say that this is a dilaton-type model, though one where the field is assumed to couple only to the electromagnetic sector of the Lagrangian. The model's dynamical equations are obtained by standard variational principles.

Specifically, the value of the fine-structure constant is related to the scalar field via $\alpha/\alpha_0= e^{2(\psi-\psi_0)}$ though the observational parameter of choice is the relative variation of $\alpha$, namely
\be
\frac{\Delta\alpha}{\alpha}(z)\equiv\frac{\alpha(z)-\alpha_0}{\alpha_0}= e^{2(\psi-\psi_0)}-1\,;
\ee
thus a negative value corresponds to a smaller value of $\alpha$ in the past. Without loss of generality we henceforth re-define the field such that $\psi_0=0$.

Assuming a flat, homogeneous and isotropic cosmology (fully in agreement with the latest cosmological data \cite{Planck}), one obtains the following Friedmann equation
\be
H^2=\frac{8\pi G}{3}\left[\rho_m(1+\omega\zeta_\alpha e^{-2\psi})+\rho_r e^{-2\psi}+\rho_\Lambda+\frac{1}{2}\omega{\dot\psi}^2  \right]\,,
\ee
with the dots denoting derivatives with respect to physical time, and the $\rho_i$ respectively denonting the matter, radiation and dark energy densities. The scalar field equation is
\be
{\ddot\psi}+3H{\dot\psi}=-2\zeta_\alpha G\rho_m e^{-2\psi} \,.
\ee
Here $\omega$ is a parameter that can be defined as $\omega\sim\hbar c/\ell^2$, where $\ell$ effectively describes the scale below which one has significant deviations from standard electromagnetism. For simplicity (and consistently with \cite{SBM,Leal}) we take $\omega\sim1$, leaving the coupling $\zeta_\alpha$ as the only phenomenological free parameter in the model. Typical values for this free parameter are discussed in some detail in \cite{SBM}, but this discussion is not strictly relevant in the present context. For our purposes, irrespective of theoretical priors, the coupling parameter $\zeta$ is taken as a free phenomenological parameter, to be constrained by observations. Note that in addition to radiation and matter the model needs a cosmological constant to match cosmological observations; the dynamical scalar field $\psi$ is subdominant in the dynamics of the universe (so we can assume the standard values of the cosmological parameters), and its only role is to drive a variation of the fine-structure constant.

In practice it is more convenient to write this equation as a function of redshift
\be\label{dyna}
\psi''+\left(\frac{d\ln{H}}{dz}-\frac{2}{1+z}\right)\psi'=-\frac{3\zeta_\alpha \Omega_m}{4\pi}(1+z)  e^{-2\psi}\left(\frac{H_0}{H}\right)^2\,,
\ee
with the primes denoting derivatives with respect to redshift. This can be straightforwardly integrated, together with the Friedmann equation, by standard numerical methods. In what follows we can safely neglect the radiation term in the Friedmann equation, since all our data only probes low redshifts, $z<5$. We will also assume a flat universe. For consistency we have kept the kinetic term of the scalar field ($1/2\, {\dot\psi}^2$), although it is easy to check that for values of the coupling consistent with the current data this term is also sub-dominant.

\begin{figure}
\includegraphics[width=3in]{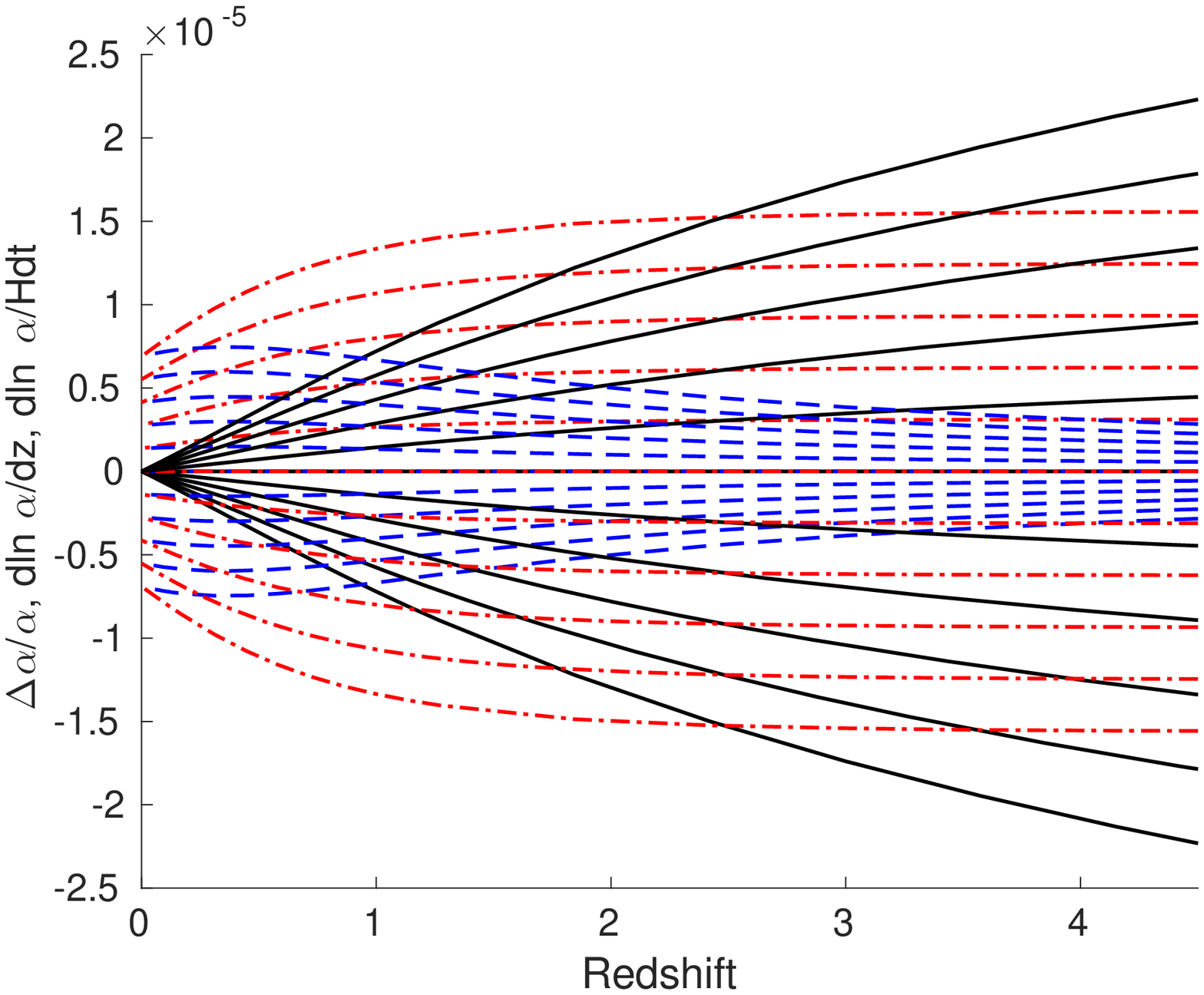}
\vskip0.2in
\includegraphics[width=3in]{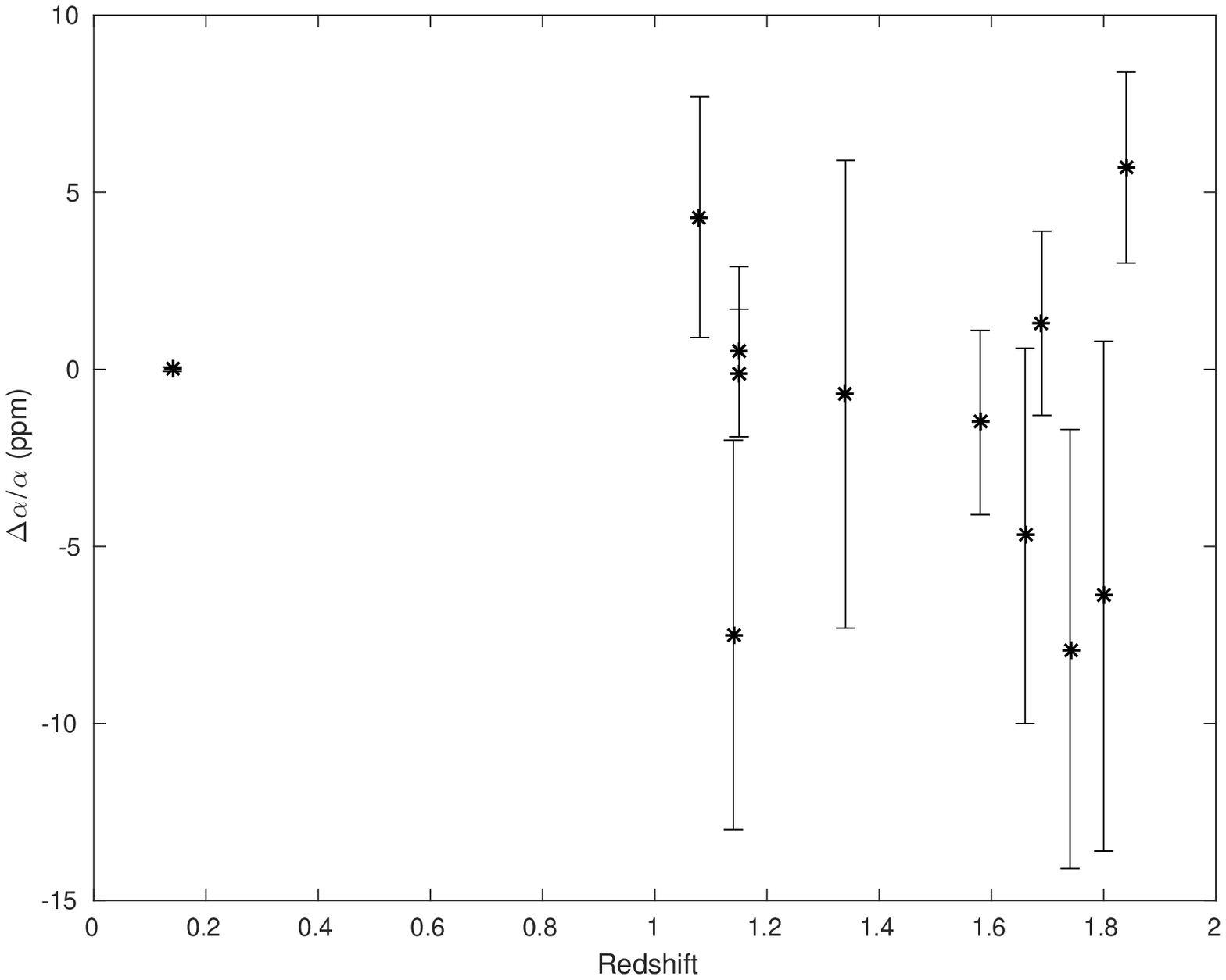}
\caption{{\bf Top panel:} Redshift evolution of $\Delta\alpha/\alpha$ (black solid lines), $\alpha'/\alpha$ (blue dashed lines) and ${\dot\alpha}/(H\alpha)$ (red dash-dotted lines) for $\Omega_m=0.3$ and couplings spanning the range $\zeta_\alpha=\pm5\times10^{-5}$. {\bf Bottom panel:} Recent measurements of $\alpha$, described in the text and listed in Table \protect\ref{table1}; see also \protect\cite{Ferreira}.}
\label{fig1}
\end{figure}

The top panel of Fig. \ref{fig1} shows the redshift evolution of $\Delta\alpha/\alpha$, the logarithmic redshift derivative $\alpha'/\alpha$ and the dimensionless time derivative ${\dot\alpha}/(H\alpha)$, for selected values of the coupling spanning the range $\zeta_\alpha=\pm5\times10^{-5}$. In particular, the current drift rate of $\alpha$, expressed in dimensionless units
\be
\left(\frac{1}{H}\frac{\dot\alpha}{\alpha}\right)_0=-\left(\frac{\alpha'}{\alpha}\right)_0=-2{\psi'}_0
\ee
will be relevant for comparison to the atomic clocks bound \cite{Rosenband}
\be
\left(\frac{1}{H}\frac{\dot\alpha}{\alpha}\right)_0=(-2.2\pm3.2)\times10^{-7}.
\ee
Moreover, in this type of model there are composition-dependent forces which lead to a WEP violation at a level quantified by the Eotvos parameter, denoted $\eta$ \cite{Will}. Specifically, for the Bekenstein-Sandvik-Barrow-Magueijo model the relation between $\eta$ and the coupling parameter is \cite{SBM}
\be
\eta_\alpha\sim3\times 10^{-9}\zeta_\alpha\,,
\ee
to be compared to current bounds coming from torsion balance experiments \cite{Torsion}
\be
\eta=(-0.7\pm1.3)\times 10^{-13}
\ee
and lunar laser ranging \cite{LLR}
\be
\eta=(-0.8\pm1.2)\times 10^{-13}\,.
\ee

As for cosmological data, we use the Union2.1 dataset of 580 Type Ia supernovas \cite{Union} as well as a set of 35 Hubble parameter measurements, of which 28 are described in the compilation of Farooq \& Ratra \cite{Farooq} while 7 more recent ones come from the work of Moresco {\it et al.} \cite{Moresco1,Moresco2}. These will provide conservative constraints on the present-day matter density, $\Omega_m$. Stronger constraints could be obtained by using for example CMB priors, but for consistency we prefer to restrict ourselves to low-redshift data.

Last but not least, we will use all the available direct astrophysical measurements of $\alpha$. We use both the Webb {\it et al.} \cite{Webb} data (a large dataset of 293 archival data measurements) and the smaller but more recent dataset of 11 dedicated measurements listed in Table \ref{table1}. The latter include the early results of the UVES Large Program for Testing Fundamental Physics \cite{LP1,LP3}, which is expected to be the one with a better control of possible systematics. Further details on this compilation can be found in \cite{Ferreira}.  Additionally we use the constraint from the Oklo natural nuclear reactor \cite{Oklo}
\begin{equation} \label{okloalpha}
\frac{\Delta\alpha}{\alpha} =(0.5\pm6.1)\times10^{-8}\,,
\end{equation}
at an effective redshift $z=0.14$; these recent $\alpha$ measurements are also plotted in the bottom panel of Fig. \ref{fig1}.

\begin{table}
\centering
\begin{tabular}{|c|c|c|c|c|}
\hline
 Object & z & ${ \Delta\alpha}/{\alpha}$ (ppm) & Spectrograph & Ref. \\
\hline\hline
3 sources & 1.08 & $4.3\pm3.4$ & HIRES & \protect\cite{Songaila} \\
\hline
HS1549$+$191 & 1.14 & $-7.5\pm5.5$ & UVES/HIRES/HDS & \protect\cite{LP3} \\
\hline
HE0515$-$441 & 1.15 & $-0.1\pm1.8$ & UVES & \protect\cite{alphaMolaro} \\
\hline
HE0515$-$441 & 1.15 & $0.5\pm2.4$ & HARPS/UVES & \protect\cite{alphaChand} \\
\hline
HS1549$+$191 & 1.34 & $-0.7\pm6.6$ & UVES/HIRES/HDS & \protect\cite{LP3} \\
\hline
HE0001$-$234 & 1.58 & $-1.5\pm2.6$ &  UVES & \protect\cite{alphaAgafonova}\\
\hline
HE1104$-$180 & 1.66 & $-4.7\pm5.3$ & HIRES & \protect\cite{Songaila} \\
\hline
HE2217$-$281 & 1.69 & $1.3\pm2.6$ &  UVES & \protect\cite{LP1}\\
\hline
HS1946$+$765 & 1.74 & $-7.9\pm6.2$ & HIRES & \protect\cite{Songaila} \\
\hline
HS1549$+$191 & 1.80 & $-6.4\pm7.2$ & UVES/HIRES/HDS & \protect\cite{LP3} \\
\hline
Q1101$-$264 & 1.84 & $5.7\pm2.7$ &  UVES & \protect\cite{alphaMolaro}\\
\hline
\end{tabular}
\caption{\label{table1}Recent dedicated measurements of $\alpha$. Listed are, respectively, the object along each line of sight, the redshift of the measurement, the measurement itself (in parts per million), the spectrograph, and the original reference. The first measurement is the weighted average from 8 absorbers in the redshift range $0.73<z<1.53$ along the lines of sight of HE1104-1805A, HS1700+6416 and HS1946+7658, reported in \cite{Songaila} without the values for individual systems. The UVES, HARPS, HIRES and HDS spectrographs are respectively in the VLT, ESO 3.6m, Keck and Subaru telescopes.}
\end{table}

\subsection{Current constraints}

We can now compare this model with the aforementioned datasets, and obtain constraints on the parameter space of the matter density $\Omega_m$ and the coupling $\zeta_\alpha$. For simplicity we will only consider flat universes and ignore the radiation density (since we are dealing with low-redshift constraints). We will further assume that the dark energy (which in this model can't be provided by the degree of freedom responsible for the $\alpha$ variation) is due to a cosmological constant.

\begin{figure}
\includegraphics[width=3in]{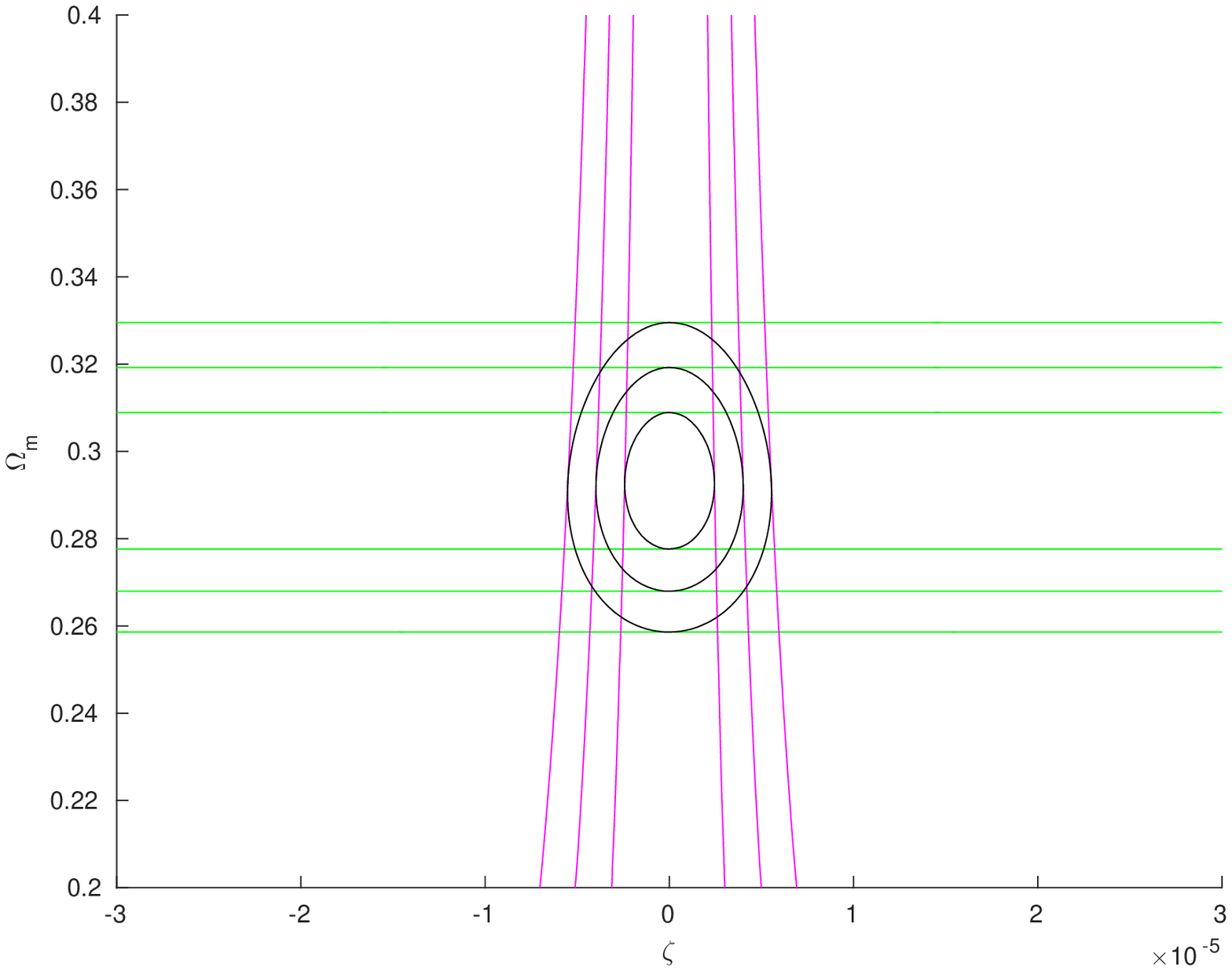}
\vskip0.2in
\includegraphics[width=3in]{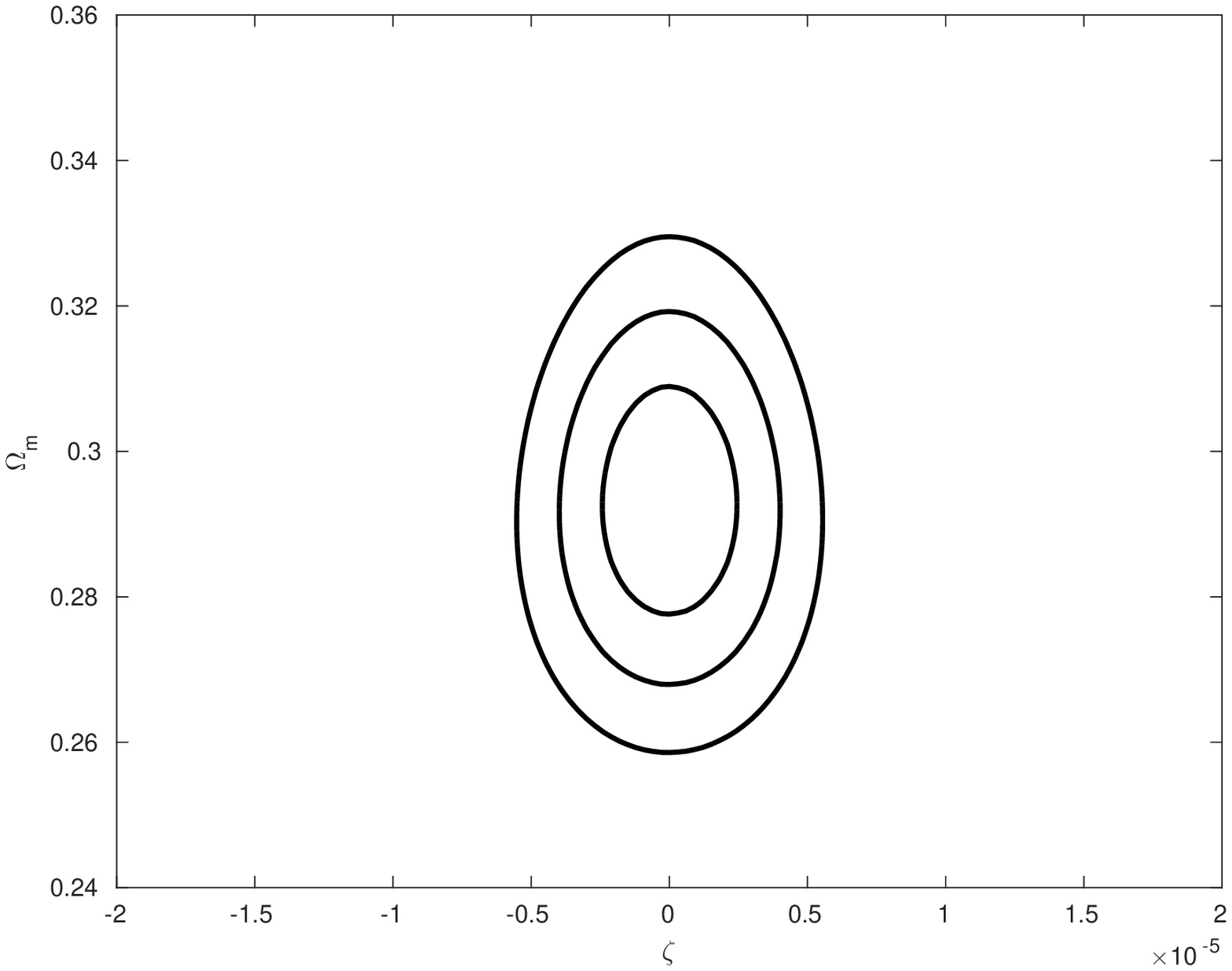}
\caption{2D likelihood contours on the $\zeta_\alpha-\Omega_m$ plane. The top panel illustrates the combination of cosmological (green horizontal contours) and astrophysical data (magenta vertical contours); the bottom panel is a zoomed version. One, two and three sigma contours are plotted in all cases. The reduced chi-square for the maximum likelihood value is $\chi^2_\nu=0.97$.}
\label{fig2}
\end{figure}

\begin{figure}
\includegraphics[width=3in]{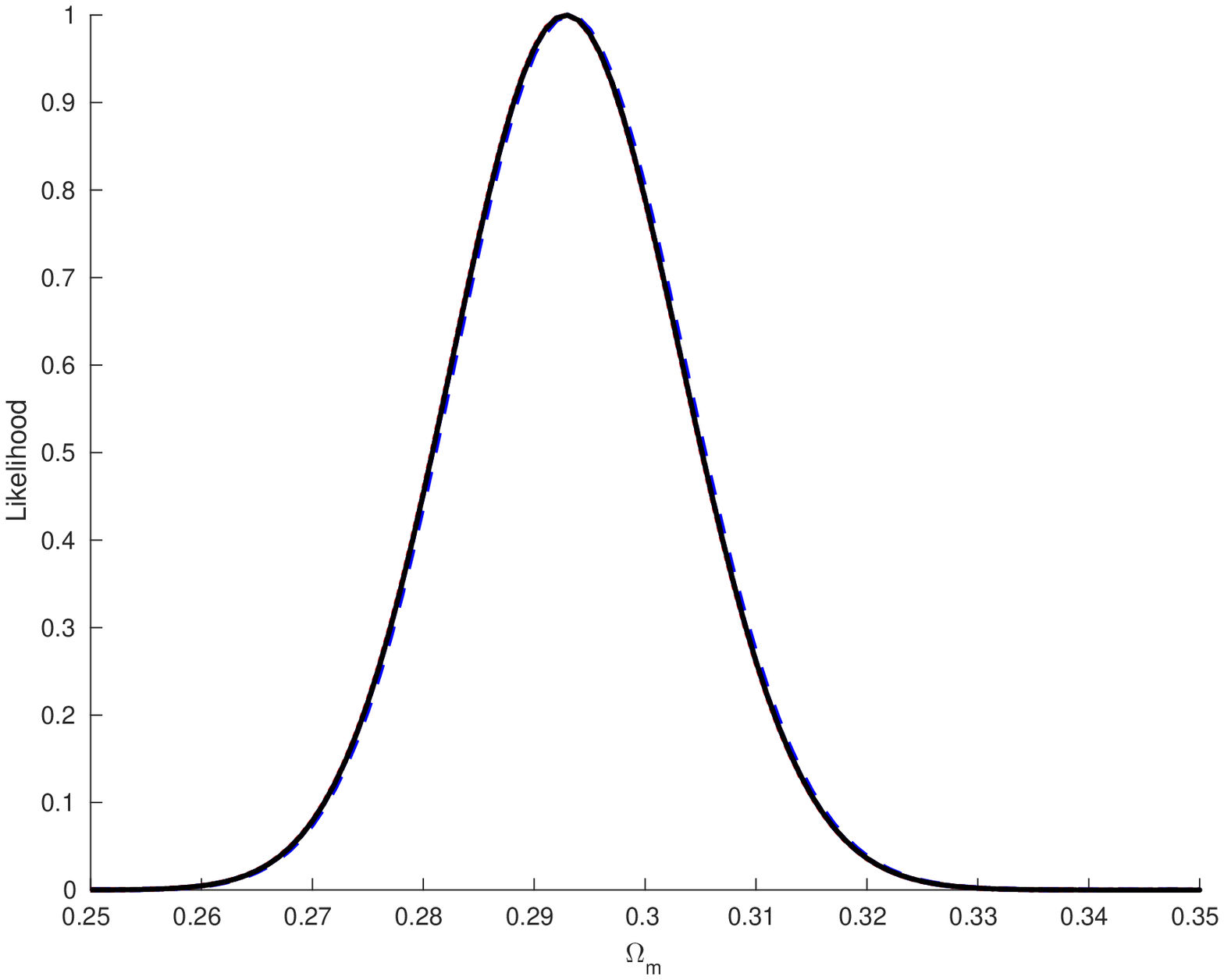}
\vskip0.2in
\includegraphics[width=3in]{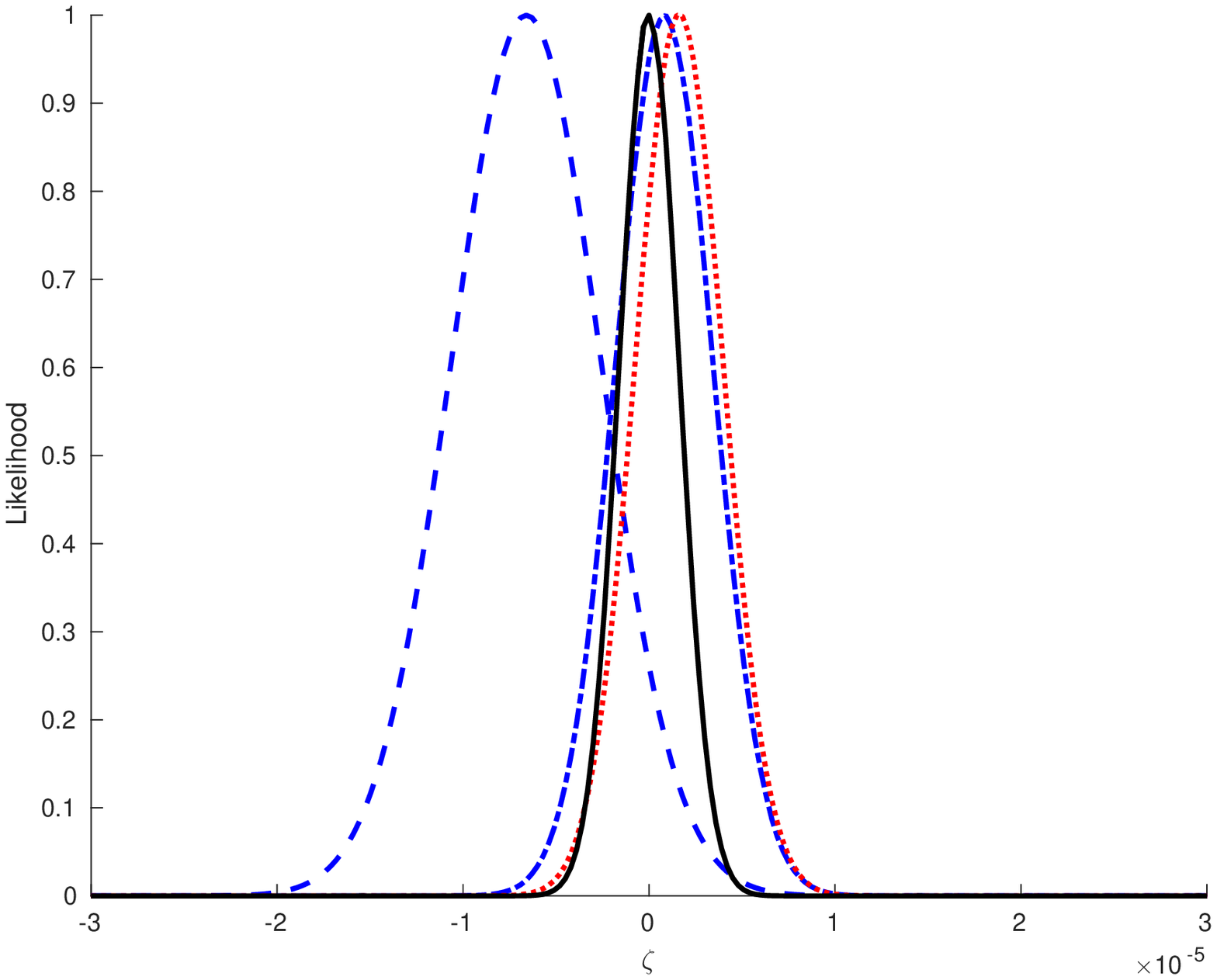}
\caption{1D likelihood for $\Omega_m$ (top) and $\zeta_\alpha$ (bottom), marginalizing over the other. In both cases the blue dashed lines correspond to the combination of cosmological and Webb {\it et al.} data, the blue dash-dotted line corresponds to the combination of cosmological, Table I and Oklo data, the red dotted line corresponds to the combination of cosmological and atomic clock data, and the black solid line corresponds to the combination of all datasets.}
\label{fig3}
\end{figure}

Our results are summarized in Figs. \ref{fig2} and \ref{fig3}. The expectation that the correlation between the two parameters is small is confirmed in Fig. \ref{fig2}. Therefore the cosmological datasets mostly fix the matter density, while the $\alpha$ measurements constrain the coupling $\zeta_\alpha$. Specifically, marginalizing over $\zeta_\alpha$ we find the following constraint
\be
\Omega_m=0.29\pm0.03\,,
\ee
at the three sigma ($99.7\%$) confidence level, which is fully compatible with other extant cosmological datasets, including the recent Planck data \cite{Planck}. As for the coupling $\zeta_\alpha$ Fig. \ref{fig3} shows that the Webb {\it et al.} dataset, combined with the cosmological data, leads to a 1.7 sigma detection of a non-zero coupling. However, the rest of the $\alpha$ datasets are compatible with the null result, with the atomic clocks bound and the recent astrophysical measurements (plus Oklo) having a comparable constraining power. This is to be contrasted with what happens in quintessence-type models, where the atomic clocks bound is more constraining than the current astrophysical measurements \cite{Pinho1,Pinho2,Pinho3}. The reason for this difference is that, whereas deep in the matter era the relative $\alpha$ variation is proportional to $\log{(1+z)}$ (just as in most dilaton-type scenarios where this behavior occurs throughout), the onset of acceleration quickly freezes the dynamics of the field and leads to comparatively smaller variations close to the present day.

For the combination of all datasets and marginalizing over $\Omega_m$, we find
\be
|\zeta_\alpha|<1.7\times10^{-6}\,,\quad 68.3\% C.L.
\ee
\be
|\zeta_\alpha|<4.8\times10^{-6}\,,\quad 99.7\% C.L.\,.
\ee
These constraints are therefore about a factor of 6 stronger than the previous analysis in \cite{Leal}, the reason being the inclusion of more recent astrophysical measurements (11 of them plus Oklo, versus only 5 measurements available to \cite{Leal}) as well as the inclusion of the atomic clocks bound \cite{Rosenband} which as not included in the earlier analysis.


Finally, we can trivially express our constraint on the coupling as a constraint on the Eotvos parameter,
\be
\eta_\alpha < 1.4\times 10^{-14}\,, \quad 99.7\% C.L.\,;
\ee
this is a stronger bound than the current local experimental limits. In other words, models in this class in agreement with the $\alpha$ constraints also satisfy current WEP bounds. However, we note that the recently launched MICROSCOPE satellite is expected to improve the sensitivity of local bounds to $\eta\sim10^{-15}$ \cite{MICROSCOPE}, thus enabling additional constraints. In parallel, the sensitivity of astrophysical bounds will also improve, as we will discuss in Sect. \ref{next}.

\section{\label{modelm}The varying $\mu$ model}

\subsection{The model and relevant data}

The Bekenstein-Barrow-Magueijo model was introduced in \cite{BM}. It is again a dilaton-type model and to a large extent analogous to the $\alpha$ model, the main difference that it describes a varying electron mass rather than a varying electric charge. Other parameters are again assumed to remain constant. Observationally, this would lead to a varying proton-to-electron mass ratio, $\mu=m_p/m_e$. (Note that \cite{BM} uses the opposite definition of $\mu$.) Consistently with the original work, we assume that the field, in this case dubbed $\phi$, driving these variations, does not significantly contribute to the Friedmann equation. We have checked that this is a good approximation given the constraints on the model.

As before we restrict ourselves to flat, homogeneous and isotropic cosmologies, neglect the radiation density (as we will again be dealing with low-redshift datasets) and assume that the dark energy is provided by a cosmological constant. In this case the electron mass is given by $m_e/m_{e0}=e^{\phi-\phi_0}$, while the observationally relevant parameter is
\be
\frac{\Delta\mu}{\mu}(z)\equiv\frac{\mu(z)-\mu_0}{\mu_0}= 1-e^{\phi-\phi_0}\,;
\ee
again a negative value corresponds to a smaller value of $\mu$ in the past. Without loss of generality we will again re-define the field such that $\phi_0=0$. In this case the dynamical equation for $\phi$, analogous to Eq. (\ref{dyna}), has the form
\be
\phi''+\left(\frac{d\ln{H}}{dz}-\frac{2}{1+z}\right)\phi'=-\zeta_\mu \Omega_m(1+z)  e^{\phi}\left(\frac{H_0}{H}\right)^2\,.
\ee
For simplicity (and in analogy with the $\alpha$ case discussed above), we have defined the dimensionless coupling $\zeta_\mu$. Written in terms of the parameters used in \cite{BM}, this is
\be
\zeta_\mu=\frac{3\Omega_b}{8\pi\mu G \omega}\left(1-\frac{f_{He}}{2}\right)\,,
\ee
where $\Omega_b$ is the present-day baryon density and $f_{He}\sim1/12$ is the Helium-4 number fraction.

\begin{figure}
\includegraphics[width=3in]{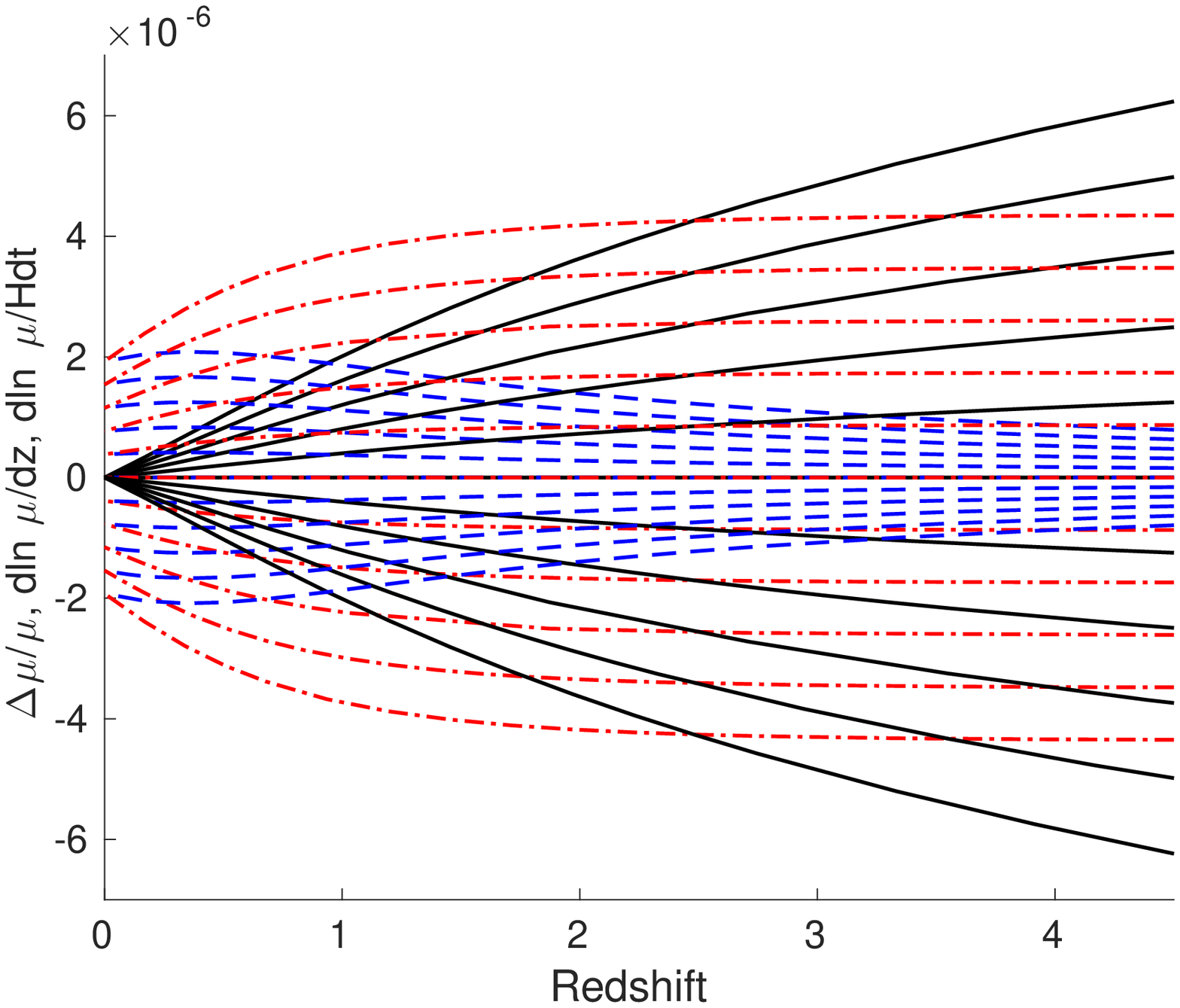}
\vskip0.2in
\includegraphics[width=3in]{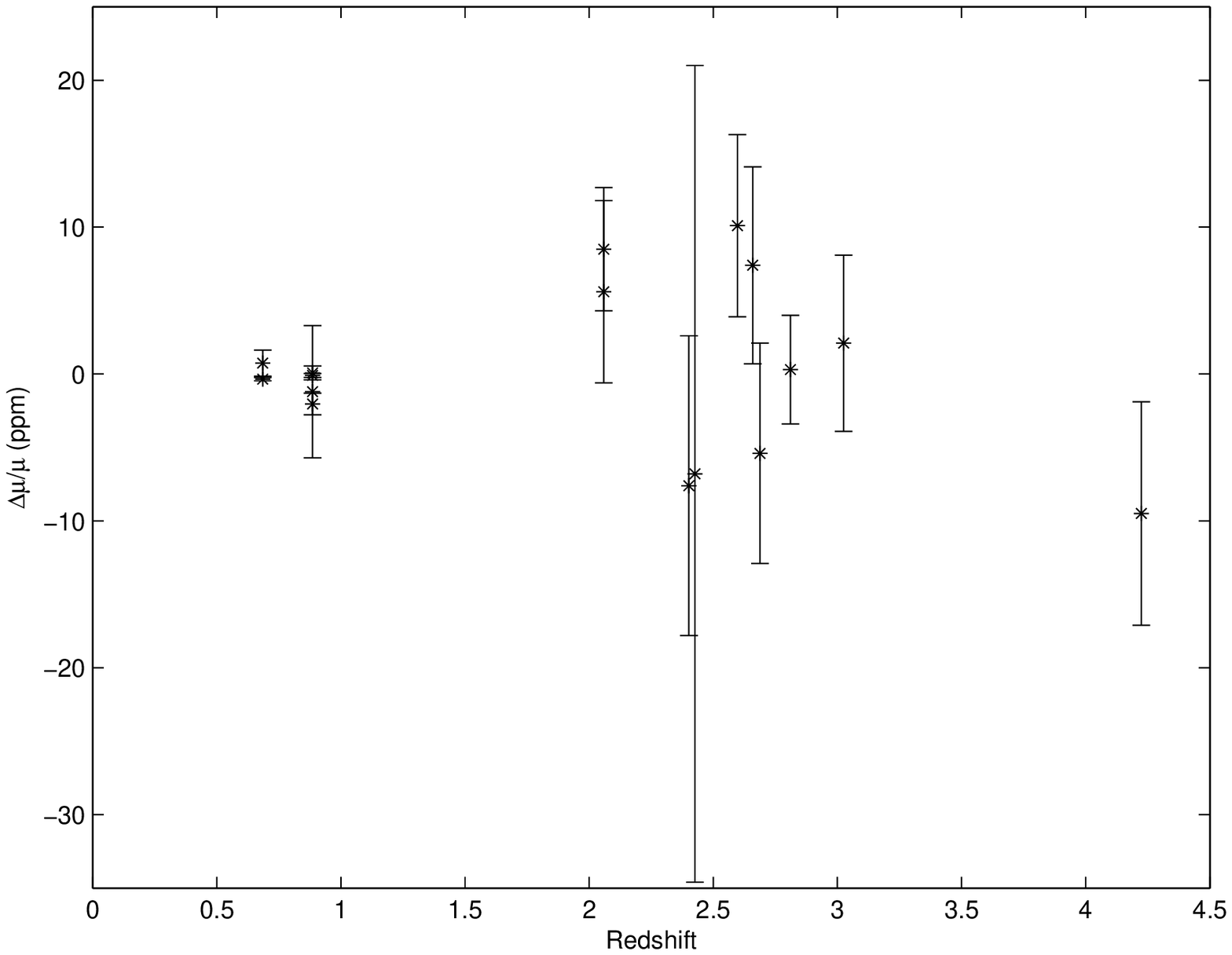}
\caption{{\bf Top panel:} Redshift evolution of $\Delta\mu/\mu$ (black solid lines), $\mu'/\mu$ (blue dashed lines) and ${\dot\mu}/(H\mu)$ (red dash-dotted lines) for $\Omega_m=0.3$ and couplings spanning the range $\zeta_\mu=\pm2\times10^{-6}$. {\bf Bottom panel:} Recent measurements of $\mu$, described in the text and listed in Table \protect\ref{table2}; see also \protect\cite{Ferreira}.}
\label{fig4}
\end{figure}

The top panel of Fig. \ref{fig4} shows the redshift evolution of $\Delta\mu/\mu$, the logarithmic redshift derivative $\mu'/\mu$ and the dimensionless time derivative ${\dot\mu}/(H\mu)$, for selected values of the coupling spanning the range $\zeta_\mu=\pm2\times10^{-6}$. We immediately see that the qualitative behavior is quite similar to that of the $\alpha$ model (c.f. the top panel of Fig. \ref{fig1}), the main quantitative difference being that, as we will presently see, the allowed values of the coupling are smaller. In this case the current drift rate of $\mu$, expressed in dimensionless units is
\be
\left(\frac{1}{H}\frac{\dot\mu}{\mu}\right)_0=-\left(\frac{\mu'}{\mu}\right)_0={\phi'}_0\,,
\ee
which is constrained by a weaker bound from local atomic clock measurements \cite{Clocks}
\be
\left(\frac{1}{H}\frac{\dot\mu}{\mu}\right)_0=(1\pm4)\times10^{-6}.
\ee
Conversely, in this case the amount of WEP violation is much stronger
\be
\eta_\mu\sim10^{-4}\zeta_\mu\,.
\ee

Naturally in this case we can use the same cosmological datasets as in the case of the $\alpha$ model. There are many fewer direct astrophysical measurements of $\mu$ than in the $\alpha$ case, although the low-redshift ones are more sensitive. A recent compilation is described in \cite{Ferreira}, and they are listed in Table \ref{table2} and plotted in the bottom panel of Fig. \ref{fig4}. Note that the measurements in the redshift range $0<z<1$ come from radio/mm observations, while those in the range $2.0<z<4.5$ come from UV/optical observations.

\begin{table}
\centering
\begin{tabular}{|c|c|c|c|c|}
\hline
 Object & z & ${\Delta\mu}/{\mu}$ & Method & Ref. \\
\hline
B0218$+$357 & 0.685 & $0.74\pm0.89$ & $NH_3$/$HCO^+$/$HCN$ & \protect\cite{Murphy2} \\
B0218$+$357 & 0.685 & $-0.35\pm0.12$ & $NH_3$/$CS$/$H_2CO$ & \protect\cite{Kanekar3} \\
\hline
PKS1830$-$211 & 0.886 & $0.08\pm0.47$ &  $NH_3$/$HC_3N$ & \protect\cite{Henkel}\\
PKS1830$-$211 & 0.886 & $-1.2\pm4.5$ &  $CH_3NH_2$ & \protect\cite{Ilyushin}\\
PKS1830$-$211 & 0.886 & $-2.04\pm0.74$ & $NH_3$ & \protect\cite{Muller}\\
PKS1830$-$211 & 0.886 & $-0.10\pm0.13$ &  $CH_3OH$ & \protect\cite{Bagdonaite2}\\
\hline
J2123$-$005 & 2.059 & $8.5\pm4.2$ & $H_2$/$HD$ (VLT)& \protect\cite{vanWeerd} \\
J2123$-$005 & 2.059 & $5.6\pm6.2$ & $H_2$/$HD$ (Keck)& \protect\cite{Malec} \\
\hline
HE0027$-$183 & 2.402 & $-7.6\pm10.2$ & $H_2$ & \protect\cite{LP2} \\
\hline
Q2348$-$011 & 2.426 & $-6.8\pm27.8$ & $H_2$ & \protect\cite{Bagdonaite} \\
\hline
Q0405$-$443 & 2.597 & $10.1\pm6.2$ & $H_2$ & \protect\cite{King} \\
\hline
J0643$-$504 & 2.659 & $7.4\pm6.7$ & $H_2$ & \protect\cite{Albornoz} \\
\hline
J1237$-$064 & 2.688 & $-5.4\pm7.5$ & $H_2$/$HD$ & \protect\cite{Dapra} \\
\hline
Q0528$-$250 & 2.811 & $0.3\pm3.7$ & $H_2$/$HD$ & \protect\cite{King2} \\
\hline
Q0347$-$383 & 3.025 & $2.1\pm6.0$ & $H_2$ & \protect\cite{Wendt} \\
\hline
J1443$+$272 & 4.224 & $-9.5\pm7.6$ & $H_2$ & \protect\cite{Bagdonaite4} \\
\hline
\end{tabular}
\caption{\label{table2}Available measurements of $\mu$. Listed are the object along each line of sight, the redshift of the measurement, the measurement itself, the molecule(s) used, and the original reference.}
\end{table}

\subsection{Current constraints}

\begin{figure}
\includegraphics[width=3in]{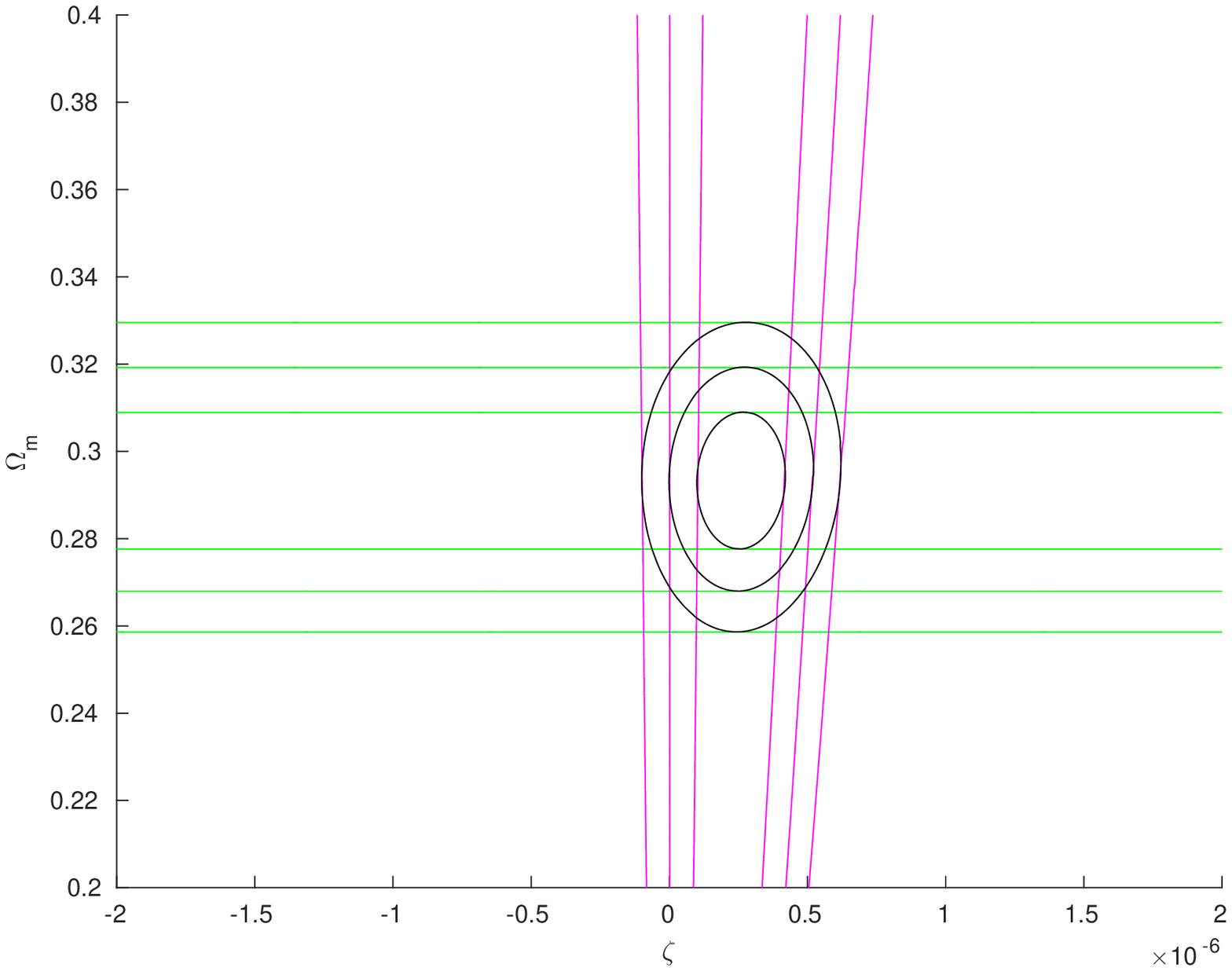}
\vskip0.2in
\includegraphics[width=3in]{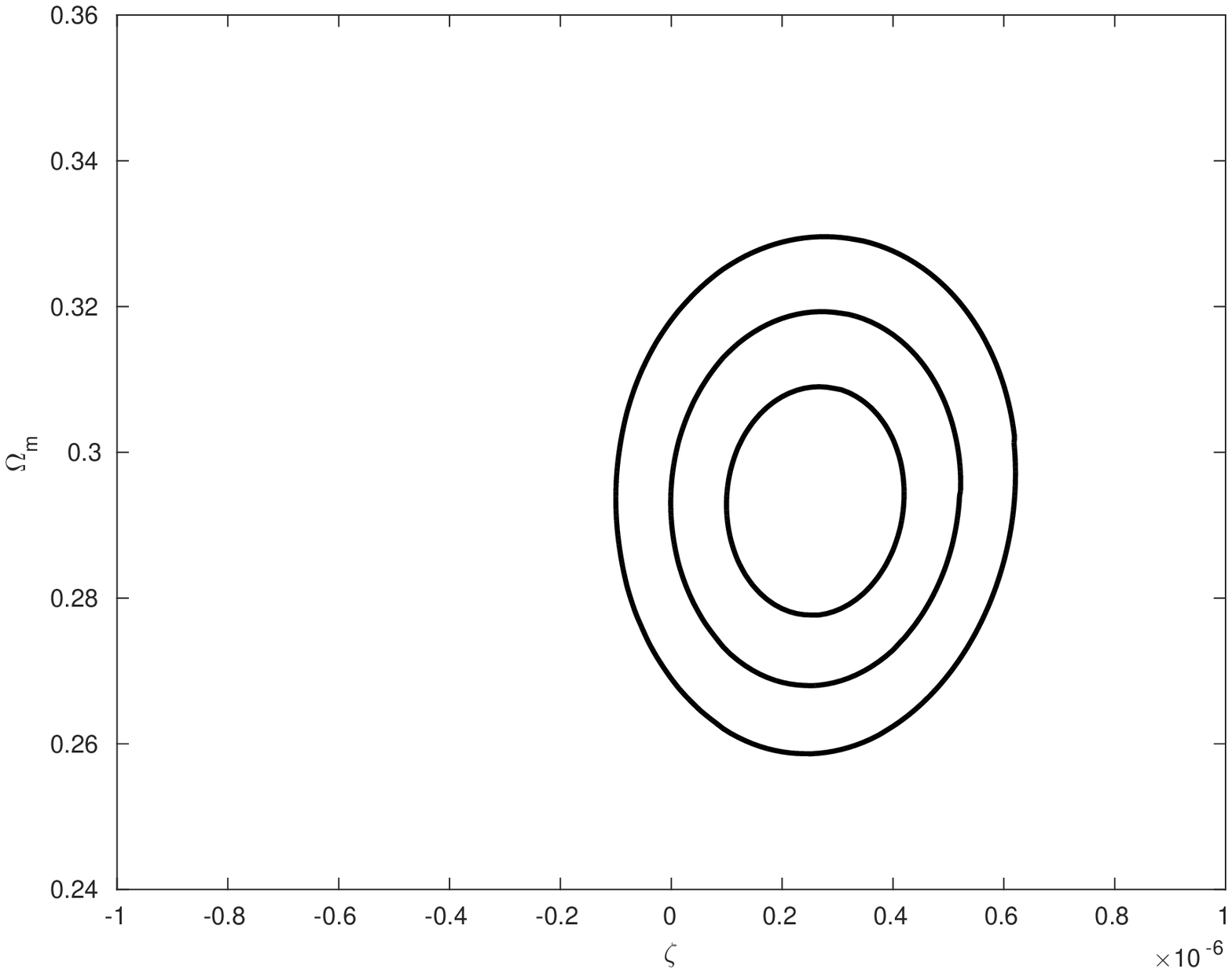}
\caption{2D likelihood contours on the $\zeta_\mu-\Omega_m$ plane. The top panel illustrates the combination of cosmological (green horizontal contours) and astrophysical (magenta vertical contours) data; the bottom panel is a zoomed version. One, two and three sigma contours are plotted in all cases. The reduced chi-square for the maximum likelihood value is $\chi^2_\nu=0.94$.}
\label{fig5}
\end{figure}

\begin{figure}
\includegraphics[width=3in]{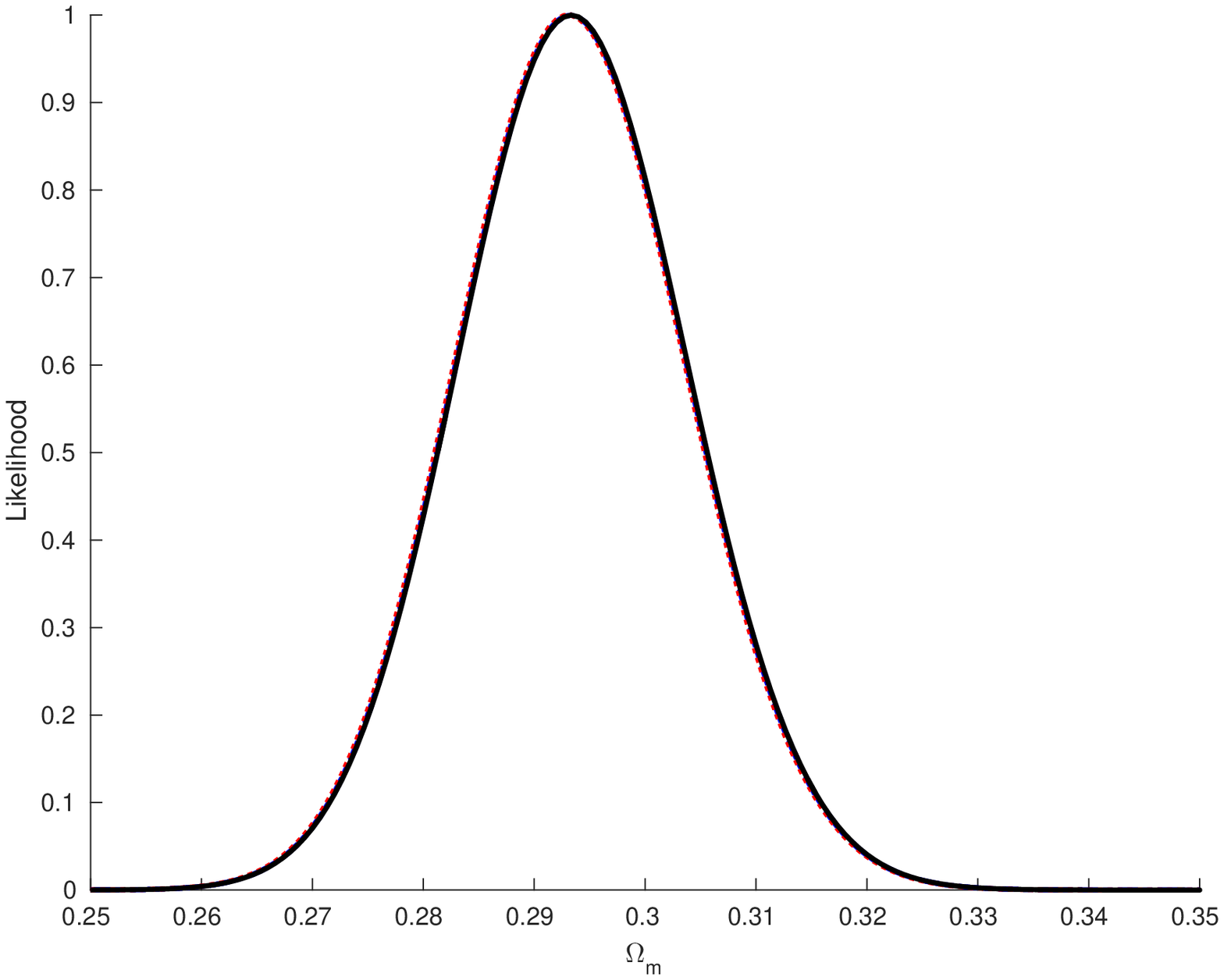}
\vskip0.2in
\includegraphics[width=3in]{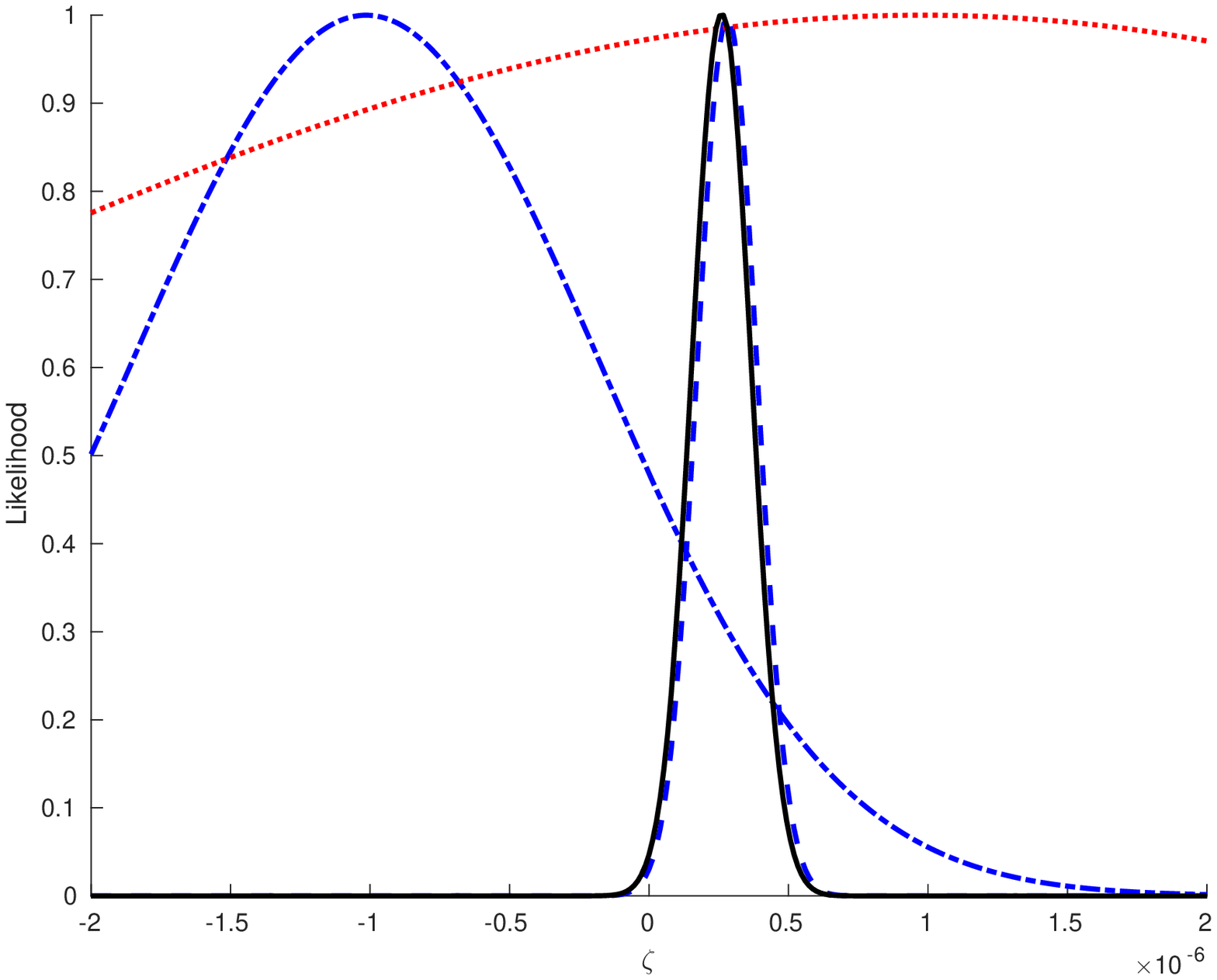}
\caption{1D likelihood for $\Omega_m$ (top) and $\zeta_\mu$ (bottom), marginalizing over the other. In both cases the blue dashed lines corresponds to the combination of cosmological and low redshift (radio/mm) $\mu$ data, the blue dash-dotted line corresponds to the combination of cosmological and high-redshift (UV/optical) data, the red dotted line corresponds to the combination of cosmological and atomic clock data, and the black solid line corresponds to the combination of all datasets.}
\label{fig6}
\end{figure}

The analysis now proceeds as in the previous chapter, and the results are summarized in Figs. \ref{fig5} and \ref{fig6}. Again the correlation between the two parameters is small, though it is different from that in the $\alpha$ model case. Marginalizing over $\zeta_\alpha$ we find the same constraint as in the previous case
\be
\Omega_m=0.29\pm0.03\,,
\ee
at the three sigma ($99.7\%$) confidence level. As for the coupling $\zeta_\mu$ Fig. \ref{fig6} shows that it is predominantly constrained by the low-redshift radio/mm astrophysical measurements of $\mu$; the constraints from the higher redshift optical/UV data are weaker (and inconsistent with the former at the 1.4 sigma level) while those from atomic clocks are even weaker. In this case, for the combination of all datasets and marginalizing over $\Omega_m$, we find
\be
\zeta_\mu=(2.7\pm1.0)\times10^{-7}\,,\quad 68.3\% C.L.
\ee
\be
\zeta_\mu=(2.7\pm3.1)\times10^{-7}\,,\quad 99.7\% C.L.\,;
\ee
this corresponds to detection of a non-zero coupling with a statistical significance of 2.5 standard deviations. However, this value of the coupling is incompatible with the bound which comes from WEP violations
\be
\zeta_\mu <4\times 10^{-9}\,, \quad 99.7\% C.L.\,;
\ee
here, unlike in the case of the $\alpha$ model, this constraint is about two orders of magnitude stronger than that coming from the $\mu$ measurements.


\section{\label{next}Forecasts for future spectrographs}

We now discuss forthcoming improvements to these constraints, expected from the new generation of high-resolution ultra-stable optical spectrographs. The first of these, ESPRESSO \cite{ESPRESSO}, will be installed at the combined Coud\'e focus of ESO's VLT in early 2017. The possibility of combining light from the four VLT unit telescopes means that ESPRESSO can therefore receive light from a 16m telescope. Thus ESPRESSO will become the instrument of choice for tests of the stability of fundamental constants until the era of the Extremely Large Telescopes, and particularly its flagship spectrograph, ELT-HIRES \cite{HIRES}.

\begin{figure}
\includegraphics[width=3in]{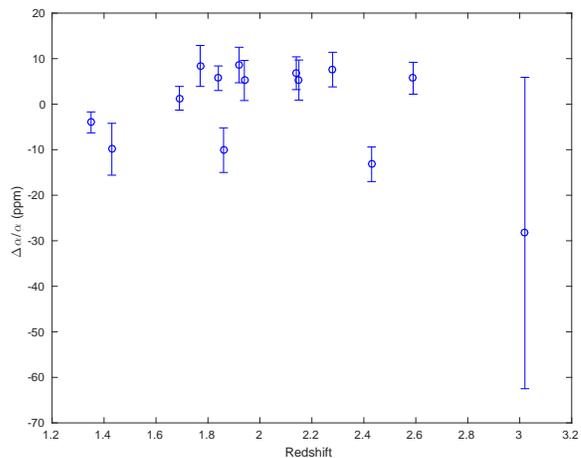}
\caption{Currently available measurements of $\Delta\alpha/\alpha$ for the list of targets selected, through the process described in \protect\cite{MSC}, for the ESPRESSO fundamental physics GTO.}
\label{fig7}
\end{figure}

A preliminary selection of the list of $\alpha$ targets to be observed during the ESPRESSO Fundamental Physics Guaranteed Time Observations (GTO) has been recently done, and the criteria and the final list are described in \cite{MSC}. This consists of 14 absorption systems, in the redshift range $1.35\le z\le 3.02$. The best currently available measurements on those systems are depicted in Fig. \ref{fig7}. Broadly speaking these are among the known systems that lead to the tightest constraints. The exception to this is the system at $z=3.02$, for which the current measurement of $\alpha$ is fairly weak; nevertheless, in this system one can also measure $\mu$ as well as the Cosmic Microwave Background temperature---see \cite{MSC} for further discussion.

For this list of ESPRESSO targets we have generated simulated measurements with the expected ESPRESSO sensitivities, assuming two different scenarios: one with a fiducial model with no $\alpha$ variation ($\zeta=0$), and the other with a fiducial model that is marginally consistent with current constraints $\zeta=5\times10^{-6}$. In either case we consider two sub-scenarios, which we call `Baseline' and `Ideal'. These are meant to represent two estimates of ESPRESSO'S actual performance and sensitivity for these measurements, with the former being conservative an the latter being somewhat more optimistic (for example, it may require additional telescope time on each target). Naturally, the actual performance of the instrument will only be known after commissioning (which will take place in 2017), but one may expect it to be somewhere between the two. Further discussion of the assumptions underlying these Baseline and Ideal scenarios can be found in \cite{MSC,Leite1,Leite2}. We also used the ESPRESSO target list for a second forecast, in this case for ELT-HIRES, by extrapolating the gains from the increased telescope collecting area \cite{Leite1,Leite2}.

\begin{figure}
\includegraphics[width=3in]{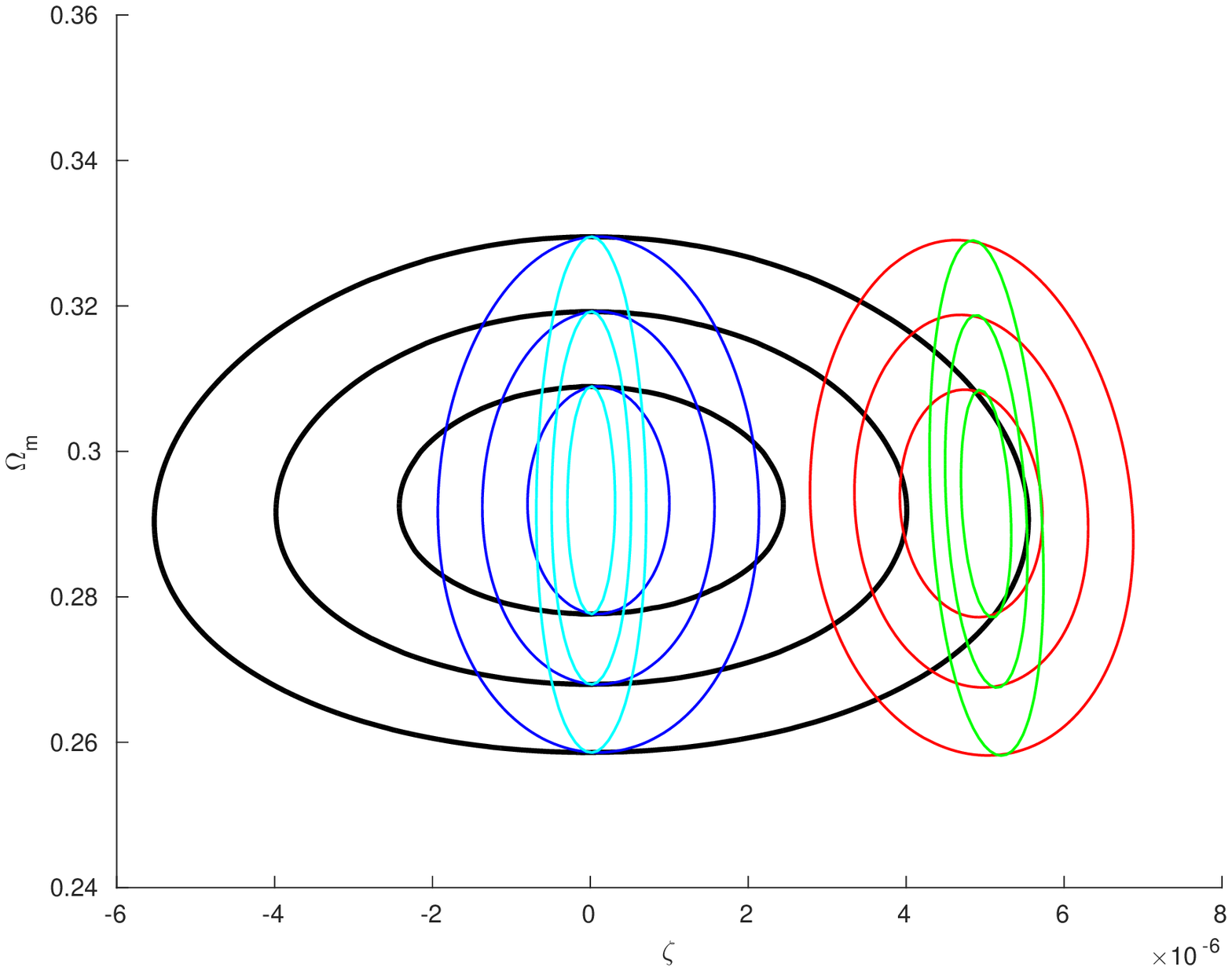}
\vskip0.2in
\includegraphics[width=3in]{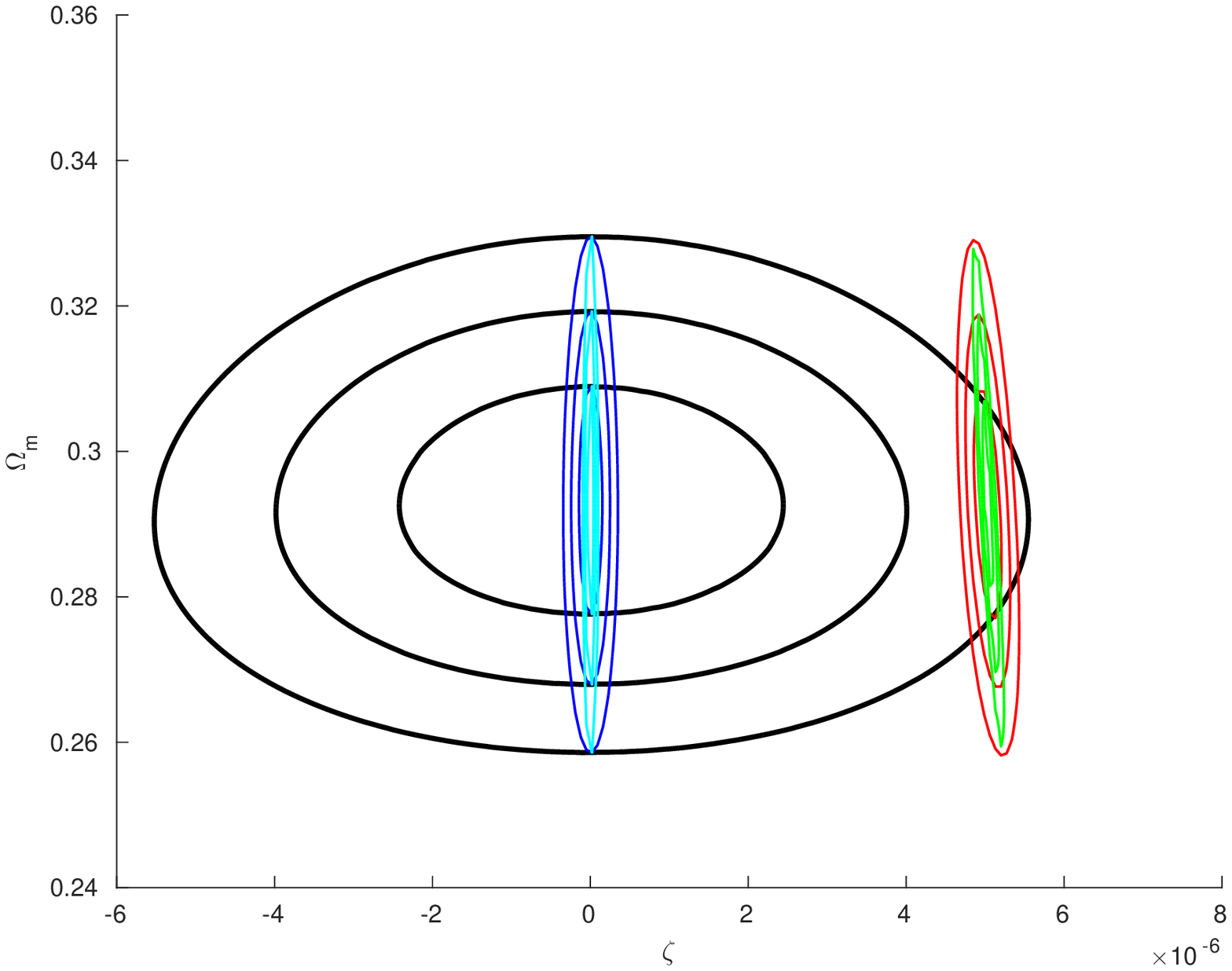}
\caption{Forecasts for the varying $\alpha$ BSBM model, from the combination of the ESPRESSO fundamental physics GTO target list \protect\cite{MSC} with the current cosmological and atomic clocks measurements used for the current data analysis. The top panel contains the forecasts for ESPRESSO itself, while the bottom one shows a forecast for the same target list observed with ELT-HIRES. In both panels the black contours correspond to the current constraints obtained in Sect. \protect\ref{modela}, the blue and cyan contours show the forecasts in the Baseline and Ideal scenarios for a fiducial model with $\zeta=0$ while the red and green contours show the forecasts in the Baseline and Ideal scenarios for a fiducial model with $\zeta=5\times10^{-6}$. One, two and three sigma contours are depicted throughout.}
\label{fig8}
\end{figure}

In all cases we generated simulated astrophysical datasets, and used them instead of the current $\alpha$ measurements (from Webb {\it et al.}, Table \ref{table1} and Oklo). Thus a dataset of 305 current measurements is replaced by one with only 14, spanning a smaller redshift range but naturally having much better precision. In repeating the analysis of Sect. \ref{modela} and forecasting the constraints on the coupling $\zeta$ we use the same cosmological datasets as well as the atomic clocks bound of Rosenband {et al.}; this is conservative assumptions since both of these datasets are also expected to improve (and in the case of the atomic clocks, the improvement may be very significant \cite{GRG}). However, our goal is to directly assess the impact of the improved astrophysical measurements.

\begin{table}
\centering
\begin{tabular}{|l|c|c|}
\hline
Dataset & $\sigma_\zeta$ ($68.3\%$ C.L.) & $\sigma_\zeta$ ($99.7\%$ C.L.) \\
\hline
Current & $1.7\times10^{-6}$ & $4.8\times10^{-6}$ \\
\hline
ESPRESSO Baseline & $6.0\times10^{-7}$ & $1.8\times10^{-6}$ \\
ESPRESSO Ideal & $2.1\times10^{-7}$ & $6.3\times10^{-7}$ \\
\hline
ELT-HIRES Baseline & $1.1\times10^{-7}$ & $3.2\times10^{-7}$ \\
ELT-HIRES Ideal & $2.3\times10^{-8}$ & $7.0\times10^{-8}$ \\
\hline
\end{tabular}
\caption{\label{table3}Current one and three sigma uncertainties on the coupling $\zeta_\alpha$ (marginalizing over $\Omega_m$) obtained in Sect. \protect\ref{modela} from current data, and the corresponding forecasts for the ESPRESSO Fundamental Physics GTO target list and the forthcoming ELT-HIRES (under the assumptions discussed in the text).}
\end{table}

The forecasts for the various cases being studied are compared with the current constraints in the $\zeta$--$\Omega_m$ plane in Fig. \ref{fig8}. These plots make it clear that even a relatively small set of only 14 measurements will lead to very significant improvements. They also show that, as was to be expected, the (small) degree of correlation between the two parameters depends on the value of the coupling. The one and three sigma forecasted uncertainties on the coupling $\zeta_\alpha$ for the various cases are plotted in Table \ref{table3}, and compared to the current ones obtained in this papers. As previously mentioned the Baseline and Ideal scenarios are intended to bracket the actual performance of ESPRESSO and ELT-HIRES (with somewhat larger uncertainties on the latter, given the earlier stage of its development). We therefore expect the ESPRESSO GTO dataset to improve constraints on the coupling in these models by a factor of 3 to 7, while ELT-HIRES can improve it by a factor of about 20 to 70 (all these factors being calculated relative to the present ones).

\section{\label{concl}Conclusions}

We have used the currently available astrophysical tests of the stability of the fine-structure constant $\alpha$ and the proton-to-electron mass ratio $\mu$, combined with local atomic clock tests and background cosmological observations to obtain improved constraints on Bekenstein-type models that lead to the variation of either. At the phenomenological level these can be seen as $\Lambda$CDM-like models with an additional dynamical degree of freedom, whose dynamics is such that it leads to the aforementioned variations without having a significant impact on the Universe's dynamics. In particular, this degree of freedom can't be responsible for the dark energy (for which a cosmological constant is still invoked), unlike the class of models recently studied in \cite{Pinho1,Pinho2,Pinho3}. In this sense these are the simplest phenomenological models allowing for $\alpha$ or $\mu$ variations, even if they are (arguably) less well motivated from a fundamental physics point of view.

While the models for $\alpha$ and $\mu$ are superficially very similar, the current experimental and observational data constrain them differently. In the $\alpha$ case the astrophysical constraints on the dimensionless coupling $\zeta_\alpha$ (which we improved by a factor of 6 compared to the previous quantitative analysis in \cite{Leal}) lead to constraints on WEP violations that are about 20 times stronger than the current direct ones (but are within the reach of the recently launched MICROSCOPE satellite \cite{MICROSCOPE}). Conversely, in the $\mu$ case, in which we have obtained the first quantitative constraints on the analogous dimensionless coupling $\zeta_\mu$ (improving upon the more qualitative analysis of \cite{BM}), the constraints are still two orders of magnitude weaker than the WEP.

We have also obtained forecasts for the target list for $\alpha$ measurements for the forthcoming ESPRESSO GTO, as well as for that same target list observed with the foreseen ELT-HIRES spectrograph. Our analysis leads to expect that constraints on the coupling $\zeta_\alpha$ should improve by a factor around 5 while for ELT-HIRES the improvement is a factor around 50. These forecasts are conservative in the sense that we assumed no improvements on cosmological, Oklo or atomic clock data (so they come from the improvements in $\alpha$ measurements alone). Moreover, further improvements could come from observing additional targets: about 300 different absorption systems have so far provided $\alpha$ measurements (and some more may be available in the future) although the 14 chosen ones are the best known ones accessible to ESPRESSO, according to the criteria discussed in \cite{MSC}. In any case, our results confirm the expectation that in the E-ELT era astrophysical tests of the stability of fundamental couplings will become a crucial part of a new generation of precision consistency tests of fundamental cosmology \cite{GRG}.

\begin{acknowledgments}

We are grateful to Ana Marta Pinho for helpful discussions on the subject of this work. This work was done in the context of project PTDC/FIS/111725/2009 (FCT, Portugal), with additional support from grant UID/FIS/04434/2013. ACL is supported by an FCT fellowship (SFRH/BD/113746/2015), under the FCT PD Program PhD::SPACE (PD/00040/2012), and by the Gulbenkian Foundation through \textit{Programa de Est\'{i}mulo \`{a} Investiga\c{c}\~{a}o 2014}, grant number 2148613525. CJM is supported by an FCT Research Professorship, contract reference IF/00064/2012, funded by FCT/MCTES (Portugal) and POPH/FSE (EC).

We thank the Galileo Galilei Institute for Theoretical Physics for the hospitality and the INFN for partial support during the completion of this work.

\end{acknowledgments}

\bibliography{bsbm}
\end{document}